\documentclass[aps,prc,twocolumn,showpacs,superscriptaddress]{revtex4}
\usepackage[english]{babel}
\usepackage{amsmath}
\usepackage{amsfonts}
\usepackage{amssymb}
\usepackage{graphicx}

\begin{document}

\title{Decay of the low-energy nuclear $^{229m}$Th isomer via atomic Rydberg states}

\author{E.~V.~Tkalya}
\email{tkalya_e@lebedev.ru}

\affiliation{P.N. Lebedev Physical Institute of the Russian
Academy of Sciences, 119991, 53 Leninskiy pr., Moscow, Russia}

\affiliation{National Research Nuclear University MEPhI, 115409,
Kashirskoe shosse 31, Moscow, Russia}

\affiliation{Nuclear Safety Institute of RAS, Bol'shaya Tulskaya
52, Moscow 115191, Russia}

\date{\today}

\begin{abstract}
In the paper, a unique process of the decay of the
$^{229m}$Th($3/2^+,8.28\pm 0.17$~eV) low energy nuclear isomer via
the internal conversion (IC) channel on Rydberg states is
considered for the first time. The Rydberg atom
$^{229m}$Th$^{+}+e^-_{Ry}$ is a unique object where IC is possible
exclusively on the Rydberg electron $e^-_{Ry}$. It is shown that
in the $^{229m}$Th$^{4+}+e^-_{Ry}$ system a) IC on the electron
states with relatively small values of the principal quantum
number $n$ and the orbital moment $l$ is practically completely
suppressed, b) IC probability, $W_{IC}$, on the
$\psi_{ns_{1/2}}({\bf{r}})$ states is proportional to
$|\psi_{ns_{1/2}}(0)|^2$ and can be related with the energy of the
hyperfine interaction of the Rydberg electron with the nucleus, c)
$W_{IC}$ decreases rapidly with the increase of $n$ in the range
from 10 to 50, and in the range $n\gtrsim 150$ $W_{IC}$ changes as
${\text{Const}}/n^3$ typical for hydrogen-like ions, d) at
$n\thickapprox 10$--30, $W_{IC}$ as a function of $l$ has a
characteristic ``knee'' between $l=3$ and $l=4$, i.e. a three
order of magnitude decrease of $W_{IC}$ due to the qualitative
change in the ratio between the centrifugal and shielding
potentials.
\end{abstract}

\maketitle

\section{Introduction}
\label{sec:Introduction}

The $^{229}$Th ground-state doublet known since the mid-70s
\cite{Kroger-76} is still a great challenge for traditional
nuclear physics. Experimental studies of the past 30 years have
shown that the energy $E_{\text{is}}$ of the first excited state
$^{229m}$Th($3/2^+,E_{\text{is}}$) probably does not exceed 10 eV
\cite{Reich-90,Helmer-94,Beck-07,Wense-16,Seiferle-17,Masuda-19,Seiferle-19}.
Such an unusually small for the nuclear scale excitation energy
leads to the appearance of decay channels of $^{229m}$Th
uncharacteristic for the nuclear spectroscopy.

Among them, it is worth mentioning the internal conversion (IC) on
the valence shells \cite{Strizhov-91} and excited atomic states
\cite{Strizhov-91,Bilous-17}, the nuclear radiation of the optical
range in dielectrics with a large band gap
\cite{Tkalya-00-JETPL,Tkalya-00-PRC}, electronic bridge
\cite{Strizhov-91,Tkalya-92-JETPL,Kalman-94,Porsev-10-PRL,Porsev-10-PRA-1+,Karpeshin-17,Muller-19,Borisyuk-19-PRC},
$\alpha$ decay of $^{229m}$Th \cite{Dykhne-96}, decay of
$^{229m}$Th in a metal via conduction electrons
\cite{Tkalya-99-JETPL}, dependence of $^{229m}$Th decay rate on
the boundary conditions \cite{Tkalya-18-PRL}, partial inversion of
the doublet levels and the decay of the ground state at the muonic
atom ($\mu^-_{1S_{1/2}}{}^{229}$Th$)^*$ \cite{Tkalya-16-PRA}, and
others.

At present, the $^{229m}$Th IC decay in atomic Th was confirmed
and the halflife of the excited state was estimated
\cite{Wense-16,Seiferle-17}, the magnetic dipole and electric
quadrupole moments of the isomeric level were measured and the
charge radius of $^{229m}$Th was determined \cite{Thielking-18}.

The importance of comprehensive studies of low-lying doublet
levels in $^{229}$Th is related to the possibility of developing
the nuclear time standard
\cite{Peik-03,Rellergert-10,Campbell-12,Peik-15}, creating a
$\gamma$-ray laser in the optical range
\cite{Tkalya-11,Tkalya-13}, measurements of the variation of
strong interaction parameter \cite{Flambaum-06,Litvinova-09} and
the fine structure constant \cite{Berengut-09}, observations of
the cooperative spontaneous emission Dicke
\cite{Dicke-54,Tkalya-11} and the M\"{o}ssbauer effect in the
optical range \cite{Tkalya-11}.

In this work, another important channel of the decay of the
$^{229m}$Th isomer is investigated --- the internal conversion on
the Rydberg electrons $e^-_{Ry}$ in the $^{229m}$Th$^{3+}$ ion
(i.e. in the $^{229m}$Th$^{4+}+e^-_{Ry}$ system). Note that
Rydberg atoms are well studied objects of modern physics (see, for
example, the reviews \cite{Gallagher-05,Sibalic-18,Lebedev-98} and
references therein). Rydberg states in the Th ions were
investigated, in particular, in the papers
\cite{Hanni-10,Keele-11,Keele-12}. Therefore, the processes
considered here in Th ions can be relatively easily studied with
the modern experimental equipment. In addition, at IC process on
the Rydberg states with large values of the principal quantum
number, the energy of the emitted electron will practically
coincide with the energy of the nuclear transition. This
simplifies the measurement of the nuclear isomer energy and the
interpretation of the experimental results.

\section{Features of IC on highly excited states in $^{229}$Th}
\label{sec:ICES}

The $^{229m}$Th isomer is the only object in which one can observe
IC on the Rydberg states in a non-ionized atom. The reason is as
follows. IC probability on the Rydberg states is vanishingly small
compared to either the usual IC, or the probability of $\gamma$
radiation at the decay of nuclear levels with excitation energy
from a few to hundreds of kiloelectronvolts. Everything is
different with $^{229m}$Th.

When one of the $(6d_{3/2})^2(7s_{1/2})^2$ valence electrons of Th
goes into an excited state, the absolute value of binding energy
of the remaining electrons increases and at some point (when a
valence electron goes into a highly excited state) it becomes
larger than the energy of the nuclear transition. For the
currently accepted value $E_{\text{is}}\approx 8.28$ eV
\cite{Seiferle-19}, the excitation of the valence $7s_{1/2}$
electron of the Th atom, say, to the 9$s_{1/2}$ state is enough to
completely stop IC process from all other levels including the
valence $6d_{3/2}$ and $7s_{1/2}$ states because in the excited
atom Th with the electron configuration
$(6d_{3/2})^2(7s_{1/2})^1(9s_{1/2})^1$ the binding energy of the
$6d_{3/2}$ electrons is approximately $-10.3$~eV, and the binding
energy of the $7s_{1/2}$ electron is $-9.6$~eV. I.e. the binding
energy of electrons at these states becomes greater than the
energy of the isomeric nuclear transition. And IC decay from the
$6d_{3/2}$ and $7s_{1/2}$ states in the excited atom Th with the
electron configuration $(6d_{3/2})^2(7s_{1/2})^1(9s_{1/2})^1$ is
forbidden by the energy conservation law. The same is true and for
the excitation of the valence $6d_{3/2}$ electron, say, to the
$9p_{1/2}$ state. In the excited atom Th with the electron
configuration $(6d_{3/2})^1(7s_{1/2})^2(9p_{1/2})^1$ the binding
energy of the $6d_{3/2}$ electron is approximately $-11.6$~eV, and
the binding energy of the $7s_{1/2}$ electrons is $-10.1$~eV.
Therefore, IC from these states is forbidden. And of course the
above is true if we excite one of the electrons to the highly
excited Rydberg state. The remaining electrons of the atomic shell
are not involved in IC.

In Th ions the ionization energy of 11.9 eV is reached already in
Th$^+$. Therefore, the decay of the $^{229m}$Th$^+$ isomer via IC
channel is possible only on excited electronic states whose
binding energy $E_{\text{b}}$ satisfies the condition
$|E_{\text{b}}|<E_{\text{is}}$. The same will take place in the
multiply charged Th ions.

The electronic bridge mentioned above can be an alternative
process (or competitor) to IC process on the Rydberg states in
$^{229}$Th. However, it is expected that it depends critically on
of the magnitude of the detuning between the energies of the
nuclear and the associated atomic transition. Therefore, the
electronic bridge will be the dominant decay channel only if
favorable conditions are met in the atom or ion for resonant
transfer of the excitation energy from the nucleus to the electron
shell.

\section{Wave functions of the Rydberg states in screened Coulomb potential}
\label{sec:Core}

Th ions can be trapped and then measured with a high accuracy in
these ion traps
\cite{Campbell-09,Campbell-11,Herrera-Sancho-13,Okhapkin-15,Thielking-18,Groot-Berning-19}.
Therefore, I will consider IC on the Rydberg electron $e^-_{Ry}$
in the $^{229m}$Th$^{3+}$ ion, which is the $^{229m}$Th$^{4+} +
e^-_{Ry}$ system (that is, the $^{229m}$Th$^{4+}$ ion with an
electron on a Rydberg state orbit). The Th$^{4+}$ ion has the
completely filled shell isoelectronic with that of the Rn atom,
which is weakly polarized by the Rydberg electron
\cite{Biemont-04,Keele-12}. A corresponding small change of the
electron density does not affect practically (after double
integration) the potential whose stems from the undisturbed
electron core of Th$^{4+}$, and consequently the Rydberg WF and
the probability of IC leave unchanged.

I consider here two cases, namely, the Rydberg states with large
and small values of the orbital moment $l$. I will call ``large''
the values of $l$ for which the Rydberg electron does not
penetrate into the region of the electron core, i.e. to the region
occupied by the shell electrons of the Th$^{4+}$ ion
\cite{Gallagher-05}.

In a hydrogen-like ion with a point nucleus of charge
$Z_{\text{nucl}}$ the potential energy of electron is
\cite{Landau-QM}
\begin{equation*}
V_{\text{H-like}}(x)=-\varepsilon_0
\frac{Z_{\text{nucl}}}{x}+V_l(x),
\end{equation*}
where $\varepsilon_0=me^4$ is the atomic unit of energy ($m$ is
the electron mass, $-e$ is the electron charge, the system of
units is $\hbar=c=1$), $x=r/a_B$, $a_B$ is the Bohr radius,
\begin{equation*}
V_l(x)=\varepsilon_0 \frac{l(l+1)}{2x^2}
\end{equation*}
is electron energy at centrifugal potential. The bound state of an
electron with the principal quantum number $n$ and the orbital
moment $l$ has in this potential two regions forbidden for the
classical motion \cite{Gallagher-05}: the first region,
$$
x > \frac{n^2}{Z}\left(1+\sqrt{1-\frac{l(l+1)}{n^2}}\right),
$$
always exists in the Coulomb potential, and the second region,
$$
0< x < \frac{n^2}{Z}\left(1-\sqrt{1-\frac{l(l+1)}{n^2}}\right),
$$
appears due to the centrifugal potential when $l\geq 1$.

In Th$^{4+}$ the characteristic size of the electron core is
approximately $2a_B$ \cite{Biemont-04}. Therefore, for each $n$
there is a value $l$, when the damped wave functions almost do not
penetrate the electron core. In this case, one can use for
calculations the electron wave functions for the potential from
the point nuclear charge \cite{Gallagher-05}.

Such functions are well known --- these are the Dirac bispinors
\cite{Akhiezer-65}. The large $g_i(x)$ and small $f_i(x)$
components of the Dirac bispinor of a bound (initial) electron
state, normalized by the condition
$\int_0^{\infty}[g_i^2(x)+f_i^2(x)]x^2dx =1$, in the Coulomb
potential of the effective point charge $Z_{\text{ion}}$ (in our
case $Z_{\text{ion}}=4$), have the form
\begin{eqnarray}
{g_i(x)\choose{}f_i(x)} & = &
-\frac{\sqrt{\Gamma(2\gamma_i+n_r+1)}}{\Gamma(2\gamma_i+1)\sqrt{n_r!}}
\sqrt{\frac{1\pm{}E_b/m}{4N(N-\kappa_i)}}\times \nonumber\\
&&(2Z_{\text{ion}}/N)^{3/2}
e^{-Z_{\text{ion}}x/N} ( 2Z_{\text{ion}}x/N)^{\gamma_i-1}\times  \nonumber\\
&& [n_r F(-n_r+1,2\gamma_i+1,2Z_{\text{ion}}x/N) \mp \nonumber\\
&& (N-\kappa_i) F(-n_r,2\gamma_i+1,2Z_{\text{ion}}x/N)]
\label{eq:WF_b},
\end{eqnarray}
where $\kappa_i=l_i(l_i+1)-j_i(j_i+1)-1/4$, $l_i$ and $j_i$ are is
the orbital and total electron momentum in the initial state,
$E_b=m/\sqrt{1+[Z_{\text{ion}}e^2/(\gamma_i+n_r)]^2}$ is the
energy of the bound electron, $n_r=n-|\kappa_i|$,
$N=\sqrt{n^2-2n_r(|\kappa_i|-\gamma_i}$,
$\gamma_i=\sqrt{\kappa_i^2 - (Z_{\text{ion}}e^2)^2}$, $F(a,b,c)$
and $\Gamma(d)$ are the confluent hypergeometric and the gamma
functions respectively \cite{Abramowitz-64}.

The wave functions of the continuous spectrum normalized at
$x\rightarrow\infty$ with the condition
$g_f(x)=i^{l_f}\sin(pa_Bx-\pi{}l_f/2)/x$ have the form
\cite{Akhiezer-65}
\begin{eqnarray}
{g_f(x)\choose{}f_f(x)} &=& \frac{e^{\pi(\xi+il_f)/2}}{x}
\frac{|\Gamma(\gamma_f+i\xi)|}{\Gamma(2\gamma_f+1)}
e^{2i\eta} \times \nonumber\\
 &&
(2pa_Bx)^{\gamma}
{1\choose{}\sqrt{(E_c-m)/(E_c+m)}} \times  \nonumber\\
 &&
{\text{Re}\choose{}\text{Im}} \left[ (\gamma_f+i\xi) e^{i\eta}
e^{-ipa_Bx} \right.
\times \nonumber\\
 &&
\left. F(\gamma_f+1+i\xi,2\gamma_f+1,2ipa_Bx)  \right]
 \label{eq:WF_c},
\end{eqnarray}
where $\gamma_f$ and $\kappa_f$ are defined through the orbital
$l_f$ and the total $j_f$ electron moments of the final state
similarly to the parameters $\gamma_i$ and $\kappa_i$ introduced
earlier, $\exp{(2i\eta)}=(\gamma_f-i\xi)/(-\kappa_f-i\xi{}m/E_c)$,
$E_c$ and $p$ are the energy and momentum of the conversion
electron satisfying the condition $E_c^2=m^2+p^2$,
$\xi=Z_{\text{ion}}e^2E_c/p$.

Electronic states with lower values of $l$ have significant
amplitudes inside the electron core. Therefore, such WFs of the
initial bound state with the electron energy $E<m$ as well as the
final state of the continuous spectrum with $E>m$ were calculated
from the Dirac equations
\begin{equation}
\left.
\begin{array}{ll}
\cfrac{dg(x)}{dx}+\cfrac{1+\kappa}{x}g(x)-
\cfrac{1}{e^2}\left(\cfrac{E}{m}+1-\cfrac{V(x)}{m}\right)f(x) =0,\\
\cfrac{df(x)}{dx}+\cfrac{1-\kappa}{x}f(x)+
\cfrac{1}{e^2}\left(\cfrac{E}{m}-1-\cfrac{V(x)}{m}\right)g(x) =0.
\end{array}
\right.
\label{eq:EqDirac}
\end{equation}
Here
\begin{equation*}
V(x)= V_{\text{nucl}}(x) + V_{\text{shell}}(x),
\end{equation*}
where $V_{\text{nucl}}(x)$ is the potential energy of electron in
potential of the unscreened nucleus, and $V_{\text{shell}}(x)$ is
the potential energy of the electron in the potential of the shell
electrons.

As $V(x)$, one can use approximate expressions for the
many-electron atoms potential from the work \cite{Flambaum-05} or
solve the Poisson equation. In the present work, the second
approach is chosen.

The electron potential energy inside and outside nucleus has been
calculated by taking the $^{229}$Th nucleus in the spherical
approximation. That is, the positive charge of the nucleus has
been uniformly distributed within a sphere of the radius $x_{R_0}
= R_0/a_B$ ($R_0 = 1.2A^{1/3}$ fm is the radius of the nucleus
with the atomic number $A$). One finds that
\begin{equation*}
V_{\text{nucl}}(x) = \left\{
\begin{array}{cl}
-\varepsilon_0
\cfrac{Z_{\text{nucl}}}{2x_{R_0}}
\left[3-\left(\cfrac{x}{x_{R_0}}\right)^2 \right] & {\text{for}}\,
0\leq{}x\leq{}x_{R_0},\\
 -\varepsilon_0 \cfrac{Z_{\text{nucl}}}{x} & {\text{for}}\, x
\geq{}x_{R_0}.
\end{array}
\right.
\label{eq:Vnucl}
\end{equation*}

\begin{figure}
 \includegraphics[angle=0,width=0.9\hsize,keepaspectratio]{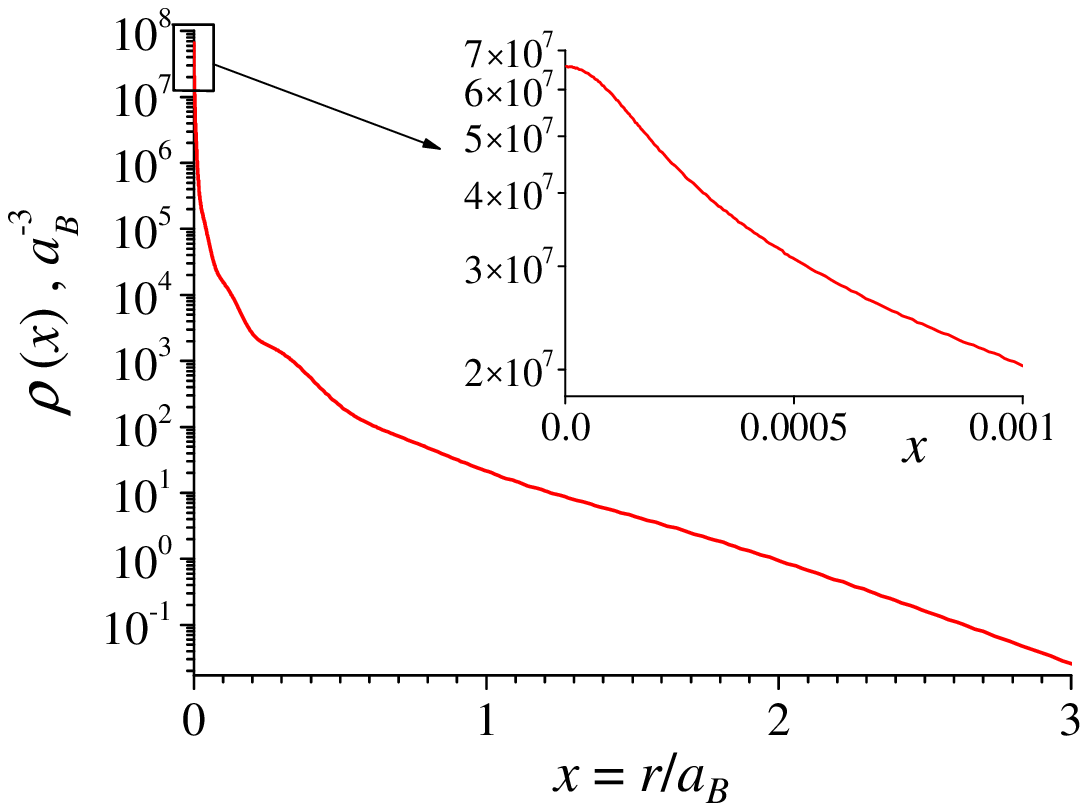}
  \caption{Electron density of the Th$^{4+}$ ion.}
  \label{fig:EDensity}
\end{figure}

The electron shell potential $V_{\text{shell}}(x)$ has been found
by solving the Poisson equation with the given electron density
$\rho_e (x)$, Fig.~\ref{fig:EDensity}. The electron density shown
in Fig.~\ref{fig:EDensity} has been obtained for Th$^{4+}$ within
the DFT theory (with three different codes
\cite{Soldatov-79,Band-79,Nikolaev-15,Nikolaev-16}) through the
self-consistent procedure taking into account the exchange and
correlation effects. The electron density for Th$^{4+}$ satisfy
the condition
$\int_0^{\infty}\rho_e(x)x^2dx=Z_{\text{nucl}}-Z_{\text{ion}}=86$.

%
%  Figure 1. Electron Density
%

Note that the implemented approach automatically takes into
account the so-called quantum defect (for more details, see in
\cite{Gallagher-05, Lebedev-98}). Here, from the very beginning I
am seeking the WFs and energies of the Rydberg states with the
quantum number $n$ as solutions of the Dirac equation
(\ref{eq:EqDirac}) in the potential $V(x)$, which takes into
account the electron shell of the ionic core. So no
renormalization of $n$ required.

\section{Internal conversion on the $ns_{1/2}$ Rydberg states}
\label{sec:ICns}

The main contribution to IC on the $ns_{1/2}$ states in the
$^{229m}$Th decay comes from the $M1$ multipole
\cite{Strizhov-91}. This is valid for IC on all electronic
$ns_{1/2}$ states including excited states
\cite{Strizhov-91,Bilous-17,Bilous-18} for the entire known range
of the reduced probabilities of the nuclear transition $0.006
\leq{} B_{\text{W.u.}}(M1,3/2^+\rightarrow 5/2^+) \leq 0.04$, $6
\leq{}B_{\text{W.u.}}(E2,3/2^+\rightarrow 5/2^+) \leq 29$
\cite{Tkalya-15-PRC,Bilous-17}. The purpose of this Section is to
estimate the probability of IC on the $ns_{1/2}$ Rydberg states
and give an approximate estimate of the magnitude of the
corresponding internal conversion coefficients (ICC). Therefore,
here I focus mainly on the discussion of the $M1$ internal
conversion.

ICC for the $E(M)L$ transition is given by the formula
\begin{eqnarray}
\alpha_{E(M)L} &=& \frac{\omega_N}{p} \frac{E_c+m}{m}
\frac{L}{L+1} \sum_{l_f,j_f}
|{\cal{F}}^{E(M)L}(l_i,j_i,l_f,j_f)|\nonumber\\
&&\times |{\cal{M}}^{E(M)L}_{if}|^2,
\label{eq:ICC_EML}
\end{eqnarray}
where functions ${\cal{F}}^{E(M)L}$ are defined by the relations
\begin{equation}
\begin{split}
{\cal{F}}^{EL}(l_i,j_i,l_f,j_f)&=(-1)^{2j_f+1} (2j_f+1)(2l_i+1)
 \times \\
&{}\left( C^{l_f0}_{L0l_i0} \right)^2 \left(
\begin{array}{ccc}
 L  & l_i & l_f \\
1/2 & j_f & j_i
\end{array}
\right)^2 ,\\
{\cal{F}}^{ML}(l_i,j_i,l_f,j_f) &=
{\cal{F}}^{EL}(l'_i,j_i,l_f,j_f) .
\end{split}
\label{eq:Fif}
\end{equation}
where $C^{l_f0}_{10l_i0}$ is the Clebsch-Gordan coefficient
followed in Eq.~(\ref{eq:Fif}) by the $6j$ symbol, $l'_i =
2j_i-l_i$. The electron matrix elements in Eq.~(\ref{eq:ICC_EML})
are
\begin{equation}
\begin{split}
{\cal{M}}^{EL}_{if} &= \int_0^{\infty}dx x^2
h_L^{(1)}(\omega_N a_B x)[g_i(x)g_f(x)+ \\
&\qquad\qquad {}f_i(x)f_f(x)],\\
{\cal{M}}^{ML}_{if} &= \cfrac{\kappa'_i-\kappa_f}{L+1}
\int_0^{\infty}dx x^2 h_L^{(1)}(\omega_N a_B x) \times \\
&\qquad\qquad {}[g_i(x)f_f(x)+f_i(x)g_f(x)].
\end{split}
\label{eq:ME}
\end{equation}
Here $h_L^{(1)}(\omega_N a_B x)$ is the Hankel function of the
first kind \cite{Abramowitz-64}, $\kappa'_i =
l'_i(l'_i+1)-j_i(j_i+1)-1/4$.

The  matrix elements (\ref{eq:ME}) were calculated by numerical
integration using analytical wave functions, Eqs.~(\ref{eq:WF_b})
and (\ref{eq:WF_c}), and numerical solutions to
Eq.~(\ref{eq:EqDirac}). Although there are cases of successful
analytical calculation of similar electronic matrix elements for
some nonrelativistic WFs of the Rydberg states
\cite{Ovsiannikov-11-PRL,Ovsiannikov-11-OS}, in our case it is
easier to use a unified approach for the both types of WFs present
in the problem.

Usually the main contribution to IC comes from the static
multipole zone near the nucleus, where the Hankel function in
Eq.~(\ref{eq:ME}) has a pole $h_l^{(1)}(z)\sim{} 1/z^{l+1}$. The
$M1$ internal conversion is most likely on the $ns_{1/2}$ states
with the maximum amplitude at the nucleus. In addition, the final
$S_{1/2}$ state of the continuous spectrum, which also has a
largest amplitude in the nucleus region, is one of the states
allowed by the selection rules for the $M1$ internal conversion on
the $ns_ {1/2}$ states. That is the internal conversion on the
$ns_{1/2}$ states give an upper estimate for IC probability on the
Rydberg levels and, accordingly, a lower bound for the nuclear
isomer lifetime.

\begin{figure}
 \includegraphics[angle=0,width=0.98\hsize,keepaspectratio]{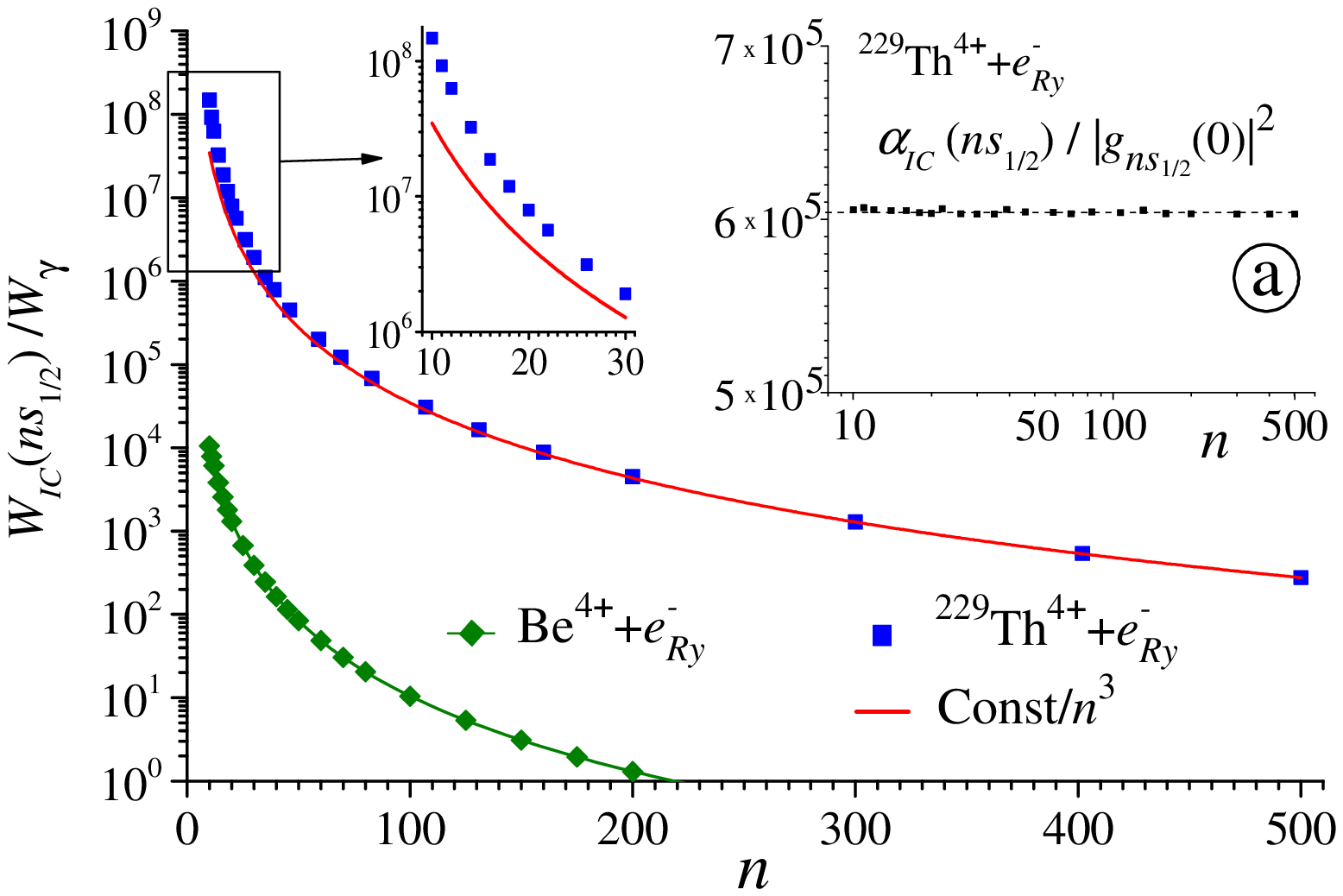}
\caption{Internal conversion coefficients for the $ns_{1/2}$
states. Blue squares are the results for
$^{229m}$Th$^{4+}+e^-_{Ry}$ (Dirac calculations), the red
connecting line is given by  $\alpha_{IC}(300)/(n/300)^3$. The
right panel inset (a) demonstrates that
$\alpha_{M1}(ns_{1/2})\propto{} g^2_{ns_{1/2}}(0)$. Green
rhombuses connected by the green line at the left lower panel are
the hydrogen-like model results for the $M1$ transition with 8.28
eV in Be$^{4+}+e^-_{Ry}$.}
  \label{fig:ICC_ns}
\end{figure}

Calculated IC coefficient $\alpha_{M1}$ as a function of the main
quantum number $n$ on the Rydberg states $ns_{1/2}$ in the
Th$^{4+}+e^-_{Ry}$ system is presented in Fig.~\ref{fig:ICC_ns}.
Up to $n = 500$, the Dirac equation (\ref{eq:EqDirac}) was solved
numerically. Using the obtained data one can extrapolate the
behavior of $\alpha_{M1}(ns_{1/2})$ to larger values of $n$. For
$n\gtrsim 150$ one finds that
$\alpha_{M1}(ns_{1/2})\sim{}{\text{Const}}/n^3$. This conclusion
is supported by the fact that in the non relativistic case for any
fixed $l_i$ the square of the modulus of the matrix element
$\langle{}f|r^{-2}|i\rangle$ for the transition from the bound
state to the continuum is proportional to $n^{-3}$
\cite{Gallagher-05}. In addition, as shown in the inset in
Fig.~\ref{fig:ICC_ns}, $\alpha_{M1}(ns_{1/2})$ is proportional to
the square of $g_{ns_{1/2}}(0)$ (which is the amplitude of WF at
$x=0$), that is, within the accuracy of our calculation
$\alpha_{M1}(ns_{1/2})/g^2_{ns_{1/2}}(0) = c_1$, where $c_1 = 6.03
\times10^5$. Taking into account that for the hydrogen-like ions
$g_{ns_{1/2}}(0) \propto (Z_{\text{ion}}/n)^{3/2}$, one obtains
again the same dependence on $n$. In particular, for the case of
the hydrogen-like Be atom (i.e. the Be$^{4+}+e^-_{Ry}$ system with
$Z_{\text{Be}}=4$) one obtains for $n \geq 10$ for the IC
coefficient the following relation
$\alpha_{M1}(ns_{1/2})=1.03\times 10^7/n^3$ (see in
Fig.~\ref{fig:ICC_ns}). This relation was calculated with the
analytical WFs (\ref{eq:WF_b})--(\ref{eq:WF_c}) for a model
nuclear $M1$ transition with the energy of 8.28 eV.

%
%  Figure 2. W_IC(n)/W_gamma
%

On the other hand, the IC probabilities for the $ns_{1/2}$ wave
functions and the other states penetrate the electron core deviate
from that for the hydrogen-like ions. For example, IC coefficients
for Th$^{4+}+e^-_{Ry}$ and Be$^{4+}+e^-_{Ry}$ shown in
Fig.~\ref{fig:ICC_ns} are very different. Even for very large $n$,
such WFs feel a charge distribution inside the electron core,
which leads to different values for electron matrix elements. In
the region of $n=10-50$ the influence of the Th$^{4+}$ electron
core near the  nucleus (where the $ns_{1/2}$ states feel a huge
charge of $^{229}$Th) is especially strong. With the increase of
$n$, the average radius of the $ns_{1/2}$ state increases as
$n^2/Z_{\text{eff}}$ (Rydberg electron is moved in effective
potential created by the effective charge $Z_{\text{eff}}$), and
the screening of the nucleus is strengthening. It leads to a rapid
decrease of $Z_{\text{eff}}$ and effective potential. As a result,
both $|g_{ns_{1/2}}(0)|^2$ and the IC probability more fast
decrease with $n$ in Th$^{4+}+e^-_{Ry}$ in comparison with
Be$^{4+}+e^-_{Ry}$ (see the deviation the IC probability from the
$n^{-3}$ law in Th$^{4+}+e^-_{Ry}$ in the left top panel inset in
Fig.~\ref{fig:ICC_ns}).

The calculation of IC probability for the $E2$ multipole gives the
same dependence of ICC as for the $M1$ case:
$\alpha_{E2}(ns_{1/2})\sim{}|g_{ns_{1/2}}(0)|^2 \propto n^{-3}$.
Thus, the well known relationship between the probabilities
$W_{IC}^{M1}(ns_{1/2}) \gg W_{IC}^{E2}(ns_{1/2})$ for IC on the
$ns_{1/2}$ states in the region of small $n$ \cite{Strizhov-91} is
preserved for large values of $n$. Using the relations
$\alpha_{M1}(ns_{1/2})\sim{}{\text{Const}}/n^3$ and
$W_{IC}^{M1}(ns_{1/2}) \gg W_{IC}^{E2}(ns_{1/2})$, one can find
the value $n$, at which the probability of the internal conversion
on the Rydberg state $ns_{1/2}$ becomes equal to the probability
of the $\gamma$ radiation upon the decay of the $^{229m}$Th
isomer. One obtains $n \simeq 3300$.

I note an interesting feature of the internal conversion on the
$ns_{1/2}$ states. The Rydberg electron in the $ns_{1/2}$ state in
the $^{229m}$Th$^{4+}+e^-_{Ry}$ or $^{229}$Th$^{4+}+e^-_{Ry}$
systems produces a magnetic field at the Th nucleus. The
interaction of the $^{229}$Th with this magnetic field in the
ground state (which magnetic moment is $\mu_{\text{gr}}=0.36$) or
isomeric state ($\mu_{\text{is}}=-0.37$ \cite{Thielking-18}) leads
to a splitting of nuclear levels. The energy of the sublevels with
the quantum number $F$ (${\bf{F}}={\bf{I}}+{\bf{s}}$, $I$ stands
for the nuclear spin, $s$ is the electron spin) is determined by
the formula for the Fermi contact interaction (see in
\cite{Abragam-61})
$$
E_F=E_{\text{int}}\frac{F(F+1)-I(I+1)-s(s+1)}{2Is},
$$
with the interaction energy
$$
E_{\text{int}}=\frac{8\pi}{3}\mu_{\text{gr(is)}}\mu_N \mu_B
g^2_i(0),
$$
where $\mu_B=e/2m$ is the Bohr magneton, $\mu_N=e/2M_p$ is the
nuclear magneton, $M_p$ is the proton mass.

Taking into account the relation $\alpha_{M1} = c_1g_i^2(0)$ valid
for $ns_{1/2}$, and expressing $g_i^2(0)$ by two different ways,
one can obtain a relation between the hyperfine splitting energy
$E_F$ and the internal conversion coefficient $\alpha_{M1}$, both
of which are measured experimentally. This can be very helpful for
additional verification of experimental results.

\section{Internal conversion on the Rydberg states
with $l\geq 1$}
\label{sec:ICl}

The internal conversion on the shells with $l\geq 1$ differs from
IC on the $ns_{1/2}$ shells. Along with the $M1$ multipole, the
$E2$ multipole can make a significant contribution to the IC
probability in $^{229m}$Th on some shells with $l\geq 1$ as it was
noted in \cite{Bilous-18}.

The contributions of the multipoles depend on the values of the
reduced probabilities of the nuclear transitions
$B_{\text{W.u.}}(M1;3/2^+\rightarrow 5/2^+)$ and
$B_{\text{W.u.}}(E2;3/2^+\rightarrow 5/2^+)$. To date, these
values have not been measured. That is why one usually uses
$B_{\text{W.u.}}(M1;3/2^+\rightarrow 5/2^+)$ and
$B_{\text{W.u.}}(E2;3/2^+\rightarrow 5/2^+)$ obtained with Alaga
rules from the available experimental data
\cite{Bemis-88,Gulda-02,Barci-03,Ruchowska-06} for the $M1$ and
$E2$ transitions between the rotation bands $3/2^+[631]$ and
$5/2^+[633]$ (see in \cite{Dykhne-98,Tkalya-15-PRC}) either
theoretical calculations \cite{Ruchowska-06,Minkov-17}.

In the first case, the average values of the reduced nuclear
transition probabilities are $B_{\text{W.u.}}(M1;3/2^+\rightarrow
5/2^+)=3.1\times 10^{-2}$ and $B_{\text{W.u.}}(E2;3/2^+\rightarrow
5/2^+)=11.7$ \cite{Tkalya-15-PRC}, and the $E2$ component makes
the main contribution to the IC process only on the $np_{3/2}$
shells: $W_{IC}(np_{3/2})/W_{IC}(ns_{1/2})\approx 2.6$. For the
$nd_{5/2}$ shells, the similar value is already less than 0.3, for
the $nd_{3/2}$ shells it is less than 0.2, and for the rest
shells, it is less than 0.1.

The calculation of the nuclear matrix elements of the low energy
isomeric transition in $^{229}$Th in the framework of the
quasiparticle-plus-phonon model was done in \cite{Ruchowska-06},
where the values $B_{\text{W.u.}}(M1;3/2^+\rightarrow 5/2^+)=
1.4\times10^{-2}$ and $B_{\text{W.u.}}(E2;3/2^+\rightarrow
5/2^+)\approx 67$ were predicted. Later, a more detailed and
modern model, taking into account practically all known aspects of
the nuclear forces, was used in \cite{Minkov-17}. This calculation
gave $B_{\text{W.u.}}(M1;3/2^+\rightarrow 5/2^+)=
0.76\times10^{-2}$ and $B_{\text{W.u.}}(E2;3/2^+\rightarrow
5/2^+)\approx 27$. With these nuclear matrix elements, the
contribution of the $E2$ multipole is an order of magnitude
greater than the contribution of the $M1$ multipole for IC on the
$np_{3/2}$ shells and is comparable with the $M1$ multipole for IC
on the $nd_{5/2}$ and $nd_{3/2}$ shells. Therefore, here I will
take into account the contribution to the IC of both $M1$ and $E2$
multipoles in Eqs.~(\ref{eq:ICC_EML})--(\ref{eq:ME}).

\begin{figure}
 \includegraphics[angle=0,width=0.98\hsize,keepaspectratio]{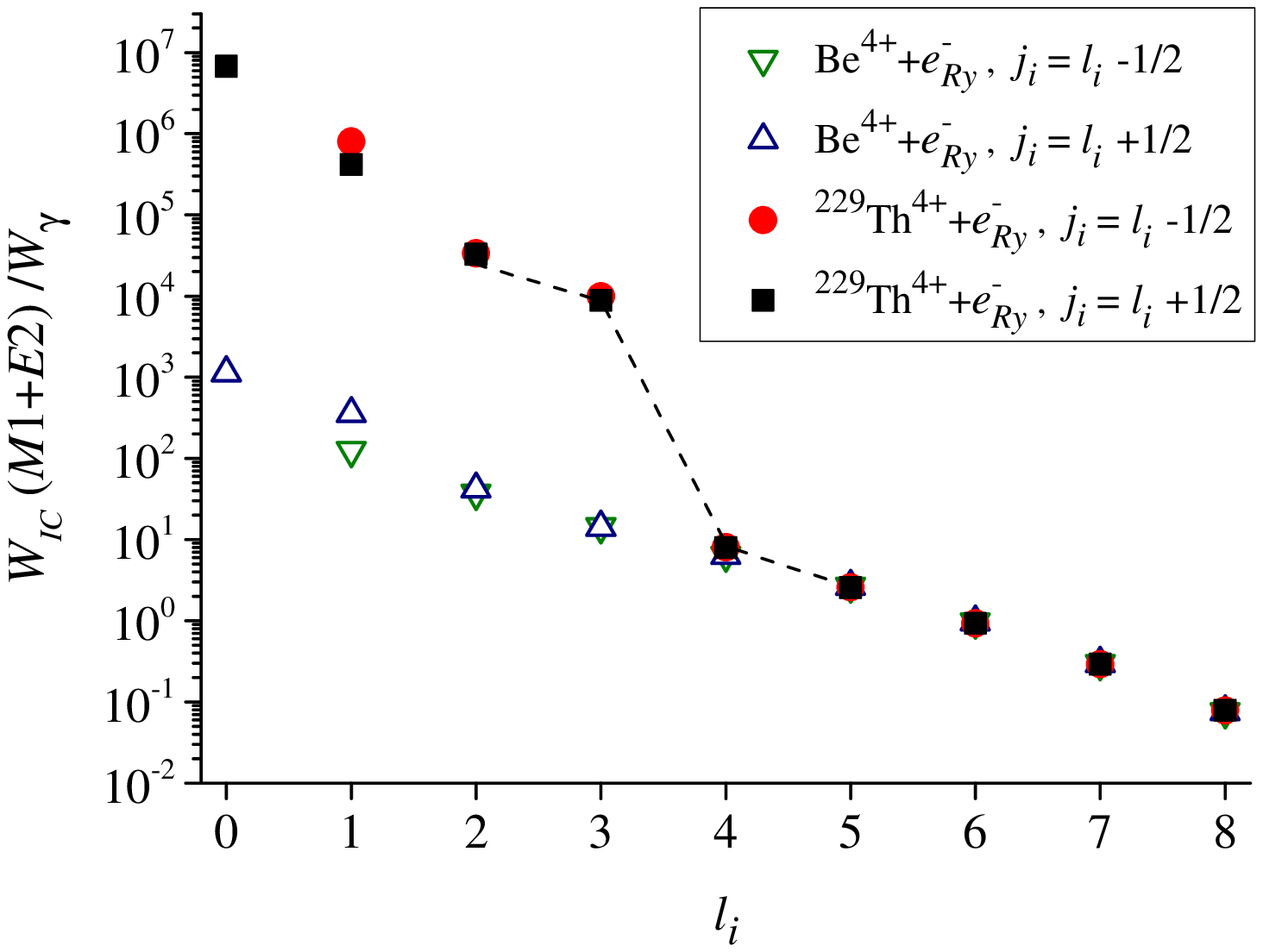}
\caption{Internal conversion coefficient as a function of the
orbital quantum number of the initial state $l_i$ for $n = 21$ in
the $^{229m}$Th$^{4+}+e^-_{Ry}$ system. The dashed line denotes
the ``knee'' --- a three order of magnitude drop of the internal
conversion process between $l_i = 3$ and $l_i = 4$. (The reduced
probabilities of the nuclear $M1$ and $E2$ transitions from work
\cite{Tkalya-15-PRC} were used for the calculations at $l_i=1$ and
2.)}
  \label{fig:ICC_l}
\end{figure}

One finds that in the $^{229m}$Th$^{4+}+e^-_{Ry}$ system the IC
probability decreases rapidly with increasing orbital angular
momentum of the initial state $l_i$. Typical dependency is shown
in Fig.~\ref{fig:ICC_l} for $n=21$. The wave functions of the
initial and final states for $0\leq{}l_i\leq 6$ were calculated by
solving the Dirac equation (\ref{eq:EqDirac}). Also, analytical
WFs, Eqs.~(\ref{eq:WF_b})--(\ref{eq:WF_c}), were used to calculate
internal conversion coefficients for $0\leq{}l_i\leq 8$ in the
hydrogen-like model for Be. One has found that the results of both
calculations lie very close for $l_i=4$ while for $l_i=5$--6 the
discrepancy is only about 1\%. Notice that for $l_i=8$--9 the
internal conversion process is practically  suppressed.

%
%  Figure 3. W_IC(l)/W_gamma
%

\begin{figure}
 \includegraphics[angle=0,width=0.98\hsize,keepaspectratio]{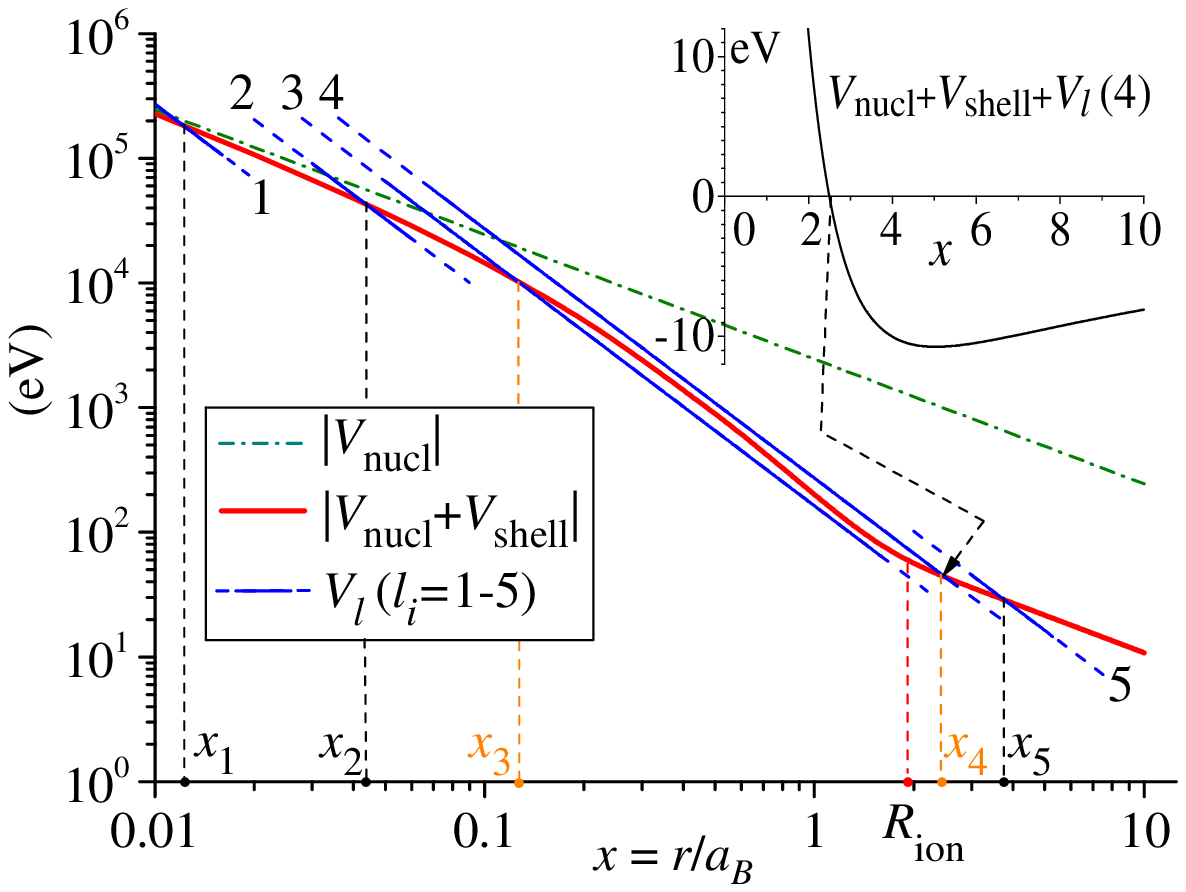}
\caption{Relations between different types of potential energy.
$V_{\text{nucl}}(x)=-\varepsilon_0 Z_{\text{Th}}/x$ is the
potential energy of the Coulomb interaction of the $^{229}$Th
nucleus and the electron, $V_{\text{nucl}}(x)+V_{\text{shell}}$ is
the energy of the Rydberg electron in the full potential (sum of
the nucleus and the electron shell term from
Eq.~(\ref{eq:EqDirac}), $V_l(x)$ is the Rydberg electron energy in
the centrifugal potential for various orbital momentum $l_i$.
$R_{\text{ion}}$ is the size of the region occupied by the
Th$^{4+}$ ion. $x_{l_i}$ ($l_i = 1$--5) are the points at which
$|V_{\text{nucl}}+V_{\text{shell}}|=V_l$ (see example for $l_i =
4$ in the right panel inset).}
  \label{fig:Potential}
\end{figure}

Earlier in the introduction I have already discussed the
mechanisms leading to the characteristic ``knee'' in
Fig.~\ref{fig:ICC_l} for $l_i=3$--4. The centrifugal potential
growing as $l_i^2$, at $l_i = 4$ exceeds the total potential of
the nucleus and the electron shell when $0\leq{}x\leq2.5$
(Fig.~\ref{fig:Potential}). Therefore, in this region in the
classical picture the motion with $l_i=4$ is forbidden, while in
the quantum case the wave functions become very small. This region
however is responsible for a significant contribution to the IC
matrix element, Eq.~(\ref{eq:ME}). (Since for the considered
values of $n$ and $l_i$ the energy of Rydberg states is very close
to zero, the boundary points of classical motion practically
coincide with the intersection points of the centrifugal potential
and total potential of the nucleus and the electron shell.) For
$l_i\leq 3$ the situation is different. The total potential
exceeds in magnitude the centrifugal potential already at
$x\gtrsim 0.1$ (see Fig.~\ref{fig:Potential}). This leads to the
formation of a potential well much closer to the nucleus than for
the $l_i = 4$ case, and to significantly larger amplitudes of the
electronic wave functions in the region responsible for a large
contribution to the integral (\ref{eq:ME}). As a result, the both
the electron matrix element of the transition and IC probability
become substantially larger.

%
%  Figure 4. Interaction Potential
%

\begin{figure}
 \includegraphics[angle=0,width=0.9\hsize,keepaspectratio]{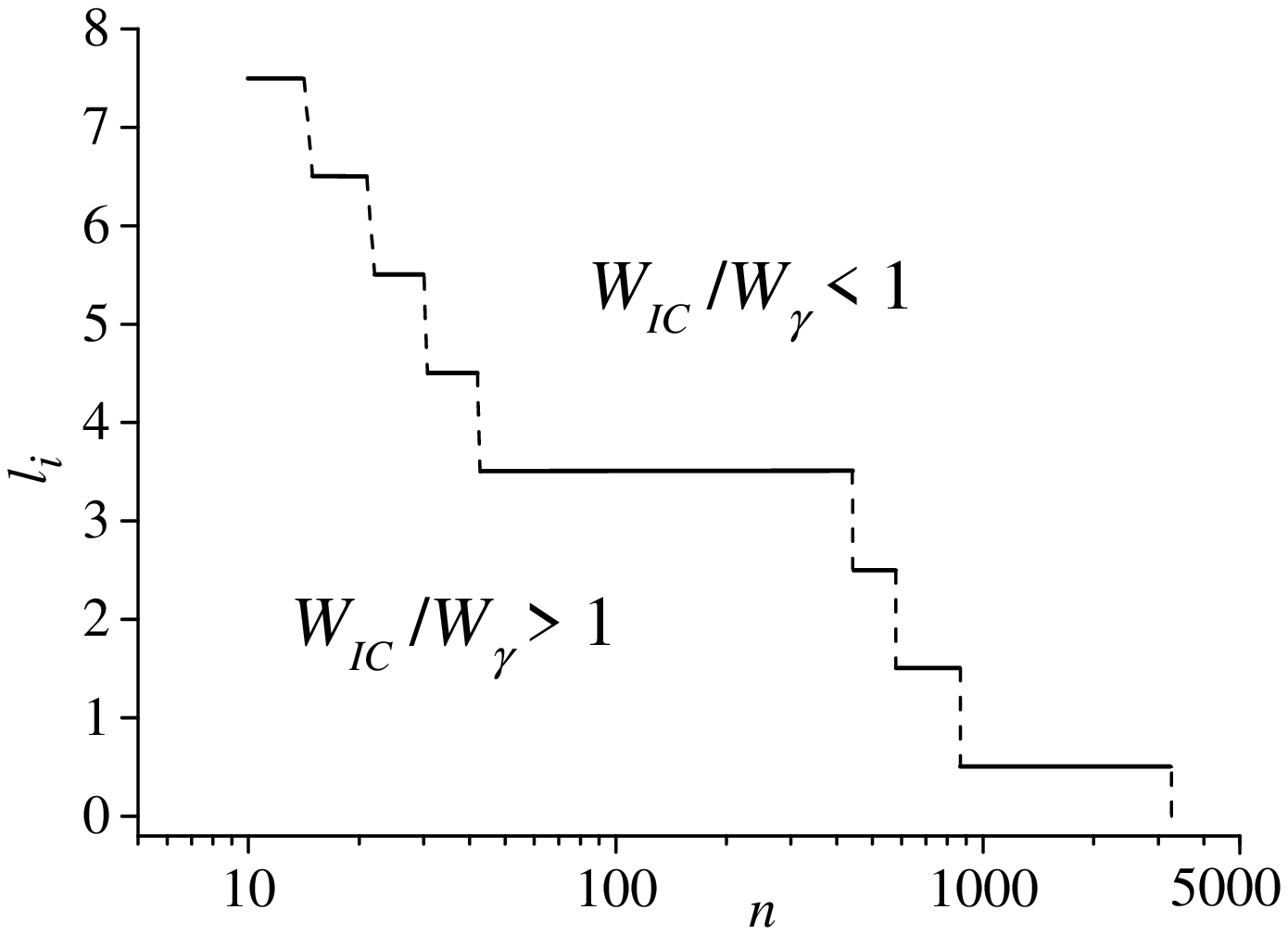}
\caption{Relative intensities of the internal conversion and the
$\gamma$ radiation versus the $n$ and $l_i$ quantum numbers.
Bottom left: the dominant channel is the internal conversion. Top
right: the dominant channel is the $\gamma$ radiation.
Solid/dashed line: boundary values of $n$ and $l_i$.}
  \label{fig:Range}
\end{figure}

The range of $n$ and $l_i$ values where IC probability in the
$^{229m}$Th$^{4+}+e^-_{Ry}$ system exceeds the probability of
$\gamma$ radiation is shown in Fig.~\ref{fig:Range}. We see that
one can control the decay of the $^{229m}$Th isomer by well known
experimental methods at easy achievable values of $n=30$--40 and
$l_i=4$--6.

%
%  Figure 5. The Range W_IC > W_gamma
%

Note in conclusion that for $l_i \geq 1$ the calculated IC
probabilities also follow the $n^{-3}$ law. This property of the
electron matrix elements (see details in \cite{Gallagher-05}) is
helpful to control the accuracy of numerical calculation of WFs
and matrix elements.

\section{Conclusion}
\label{sec:Conclusion}

Summing up, I have shown the possibility of observing
experimentally a unique process --- the decay of the low energy
isomeric nuclear level in $^{229}$Th via the channel of the
internal conversion on the Rydberg states with a substantial (many
orders of magnitude) increase of the lifetime of the $^{229m}$Th
isomer.

The author thanks Prof. N.~N.~Kolachevsky and Prof. V.~S.~Lebedev
for helpful discussions. This research was supported by a grant of
the Russian Science Foundation (Project No 19-72-30014).

\end{document}